# An Efficient Method for Mining Event-Related Potential Patterns

Seyed Aliakbar Mousavi[1], Muhammad Rafie Hj Arshad[2], Hasimah Hj Mohamed[3] and Saleh Ali Alomari[4]

[1, 2,3,4] School of Computer Sciences, Universiti Sains Malaysia
Pulau Penang 11800, Malaysia

**Abstract**
In the present paper, we propose a Neuroelectromagnetic Ontology Framework (NOF) for mining Event-related Potentials (ERP) patterns as well as the process. The aim for this research is to develop an infrastructure for mining, analysis and sharing the ERP domain ontologies. The outcome of this research is a Neuroelectromagnetic knowledge-based system. The framework has 5 stages: 1) Data pre-processing and preparation; 2) Data mining application; 3) Rule Comparison and Evaluation; 4) Association rules Post-processing 5) Domain Ontologies. In 5$^{th}$ stage a new set of hidden rules can be discovered base on comparing association rules by domain ontologies and expert rules.

## 1. Introduction

The Electroencephalographics (EEG) [1] are electric signals and Magnetoencephalographys (MEG) [2] are magnetic fields produced by electrical currents in cortex. MEG and EEG are used to record brain activities. There are other methods of imaging brain signals such as functional Magnetic Resonance (fMRI) and Imaging and Position Emission Tomography (PET). In comparison EEG/MEG method have two main advantages; direct measure of neuronal activity and best temporal resolution by milliseconds. Unlike fMRI and PET, all sensing and processing will take place within few hundred milliseconds so it gives a very good representation of time course of brain activity to control EEG signals in time. Fig.1 is an example of EEG recorded data that channels are shown vertically and time courses horizontally. EEG patterns are recorded using a set of sensors called Net which is put all around head and well connected to scalp. Therefore the electrical activities close to sensors (channels) are captured even though they are very small. Fig.2 is an example of MEG data.

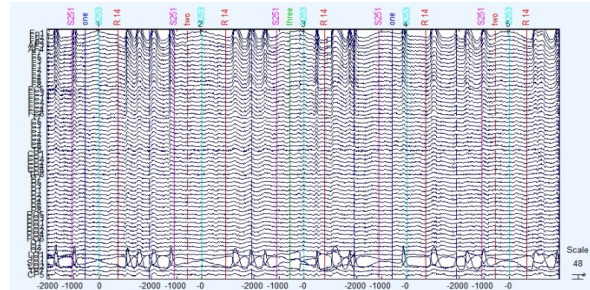

Fig.1. EEG recorded data that shows the channels vertically, time courses horizontally and events on top

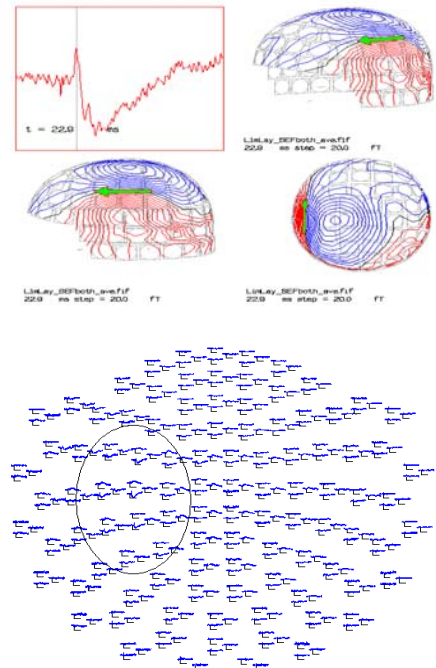

Fig.2 is MEG data that is mapped on scalp.

As shown in Figure 3 the event-related potentials (ERP) [3] are physiological patterns created by averaging segments of EEG/MEG, time-locking to event of interest (e.g., onset of a visual or auditory stimulus) such as internal and external stimulus or response onset. This notion has been generalized under the term event-related responses to include triggered events. In the latter case the recorded signal portrays the neural activity related with cognitive processes like performing a given memory task or during decision making. In both cases the target activity is accompanied by (and occasionally modulated by or even interrelated with) the ubiquitous spontaneous activity of the brain [12]. Signals that are not event-related tend towards zero as the number of averaged trials in experiment increase. In this way, averaging increases the signal-to-noise ratio (SNR) and provides measures of electrical activities that are specifically linked to stimulus processing [11]. ERP is spatial, overlapping, and temporal, so it`s difficult to extract the features by normal tools and methods. In the other hand, the data contains a lot of noise (e.g. Artifacts) and other brain activities which superpose the target signals. ERP provides the measures of brain activities that are related to stimulus processing. The ERP is characterized by the time course, polarity and scalp`s topography. It might be early or late in the time respond, either positive or negative and distributed over scalp topography.

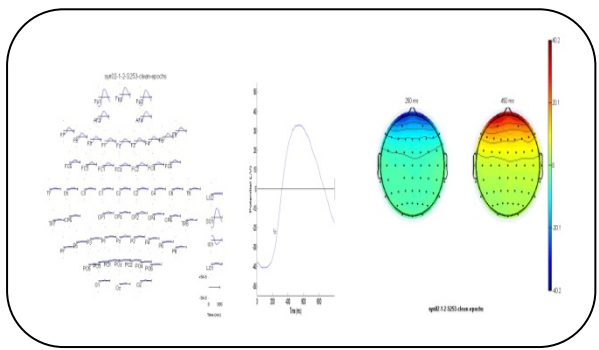

In Fig.3 there is an example of ERP with its properties from EEGlab. [7] The "P300" component (Signal) which was extracted from the superposed (Overlapped) data using Independent Components Analysis (ICA) has a peak latency of approximately 300 to 500ms and is positive over frontal areas of the scalp.

## 2.0 Related Research

Human Brain is continuously dispersing huge amount of electromagnetic waves. Brain`s electromagnetic fields are actual carriers of conscious experience [17]. Brain waves are unprocessed data which contains knowledge about brain functions. They act as input in knowledge discovery process and data mining. The brain wave patterns are important in computational neuroscience and neuro-Informatic research. The reason those patterns are retrieved is because of usefulness in recognizing brain disorders; such as by identifying the type of seizure occurring in the brain, and in diagnosing epilepsy. We can also examine the problem of dementia which is difficult to diagnose by ordinary clinical diagnosis. Furthermore, those patterns provide a practical method of identifying patients who suffer from mental health problems or are in a comatose condition. In the clinical field, EEG is applied in the neurophysiological area and also being studied in an attempt to provide a new way of communication between the human brain and computers by reading the signal extracted from the brain via the computer interface. The extracted knowledge from EEG and MEG patterns leads neuroscientist and neurosurgeons to find the patterns associated with Cortex in-functionality, Neuron illness, brain capability analysis, [4] and responses to outside stimulus in different situations. The knowledge about brain patterns can be used in commercial marketing and decision making known as Neural-Economy to increase sale and find the profitable customers.

Researches in Computational Neuroscience are depended on data and in-order to process data, there is an increasing need for advance tools and concepts. The concept of building a framework to do knowledge extraction and representation has been proposed by Dou, D., Frishkoff, G., Rong, J., Frank, R., Malony, A., & Tucker, D. in 2007 as a project called Neuro Electro Magnetic Ontologies (NEMO). In NEMO project the EEG brain waves are fed to data mining process to extracted rules and domain ontology. NEMO project along with previous researches have built frameworks and determined methods. However most of those researches have same problem which is wrong classification of ERP patterns. Those patterns split across wrong classes. Dou, D suggested a further refinement of ERP during decomposition. The reason is that ERP is spatial, temporal and with high dimension at each time point. Many parts of the brain are simultaneously active and that would overlap the other patterns to target signals which would cause uncertainty in data mining progress. He also advised to employ spatial and dense merits that capture temporal and spatial attributes more accurately, and it may reduce the misallocation of patterns variance. He suggested evaluating the machine generated rules with "Gold Standards" which has been established before by experts in ERP domain.

The main challenge which is mentioned in most of ERP researches is the establishment of robust computational analysis methods such as data decomposition (Pre-processing) and data mining to correctly interpret (classifying) ERP patterns and relate them to specific brain functions [5]. M. Sabeti, S.D. Katebi and R. Boostani [19] suggested a new approach for classification of EEG signals in schizophrenic patients. They used autoregressive model parameters, band power and fractal dimension features for classification. They employ both linear discriminant analysis (LDA) and adaptive boosting (Adaboost) [19] to classify the reduced feature set. They attained a classification with high accuracy by LDA and Adaboost, respectively. M. Sabeti and R. Boostani [19] have also tried to determine the more informative channels in EEG experiments of 20 schizophrenic patients. Gwen A. Frishkoff, RobertM. Frank, Jiawei Rong [9], suggested an automatic classification and labeling of brain electromagnetic patterns. They designed an otology-base system to classify EEG ERP patterns. They focus on specification of rules and concepts that capture expert knowledge of ERP patterns in word recognition. Their contribution is implementation of rules in an automated data processing and labeling stream along with data mining techniques that combine top-down (knowledge-driven) with bottom-up (data-driven) method to refine the ERP machine generated rules. C.Davatzikos, K. Ruparel, Y. Fan and D.G. Shen [22] used data mining machine learning to classify spatial fMRI patterns in application of lie detection. Classified, high-dimensional and non-linear pattern in functional magnetic resonance (fMRI) images is used to discover the spatial patterns of brain activities associated with lie and truth.

The methodology in most of related publications consisted of 4 phases. First is data collection and acquisition; in this part the signals are acquired from brain and collected by tools/softwares such as Neuroscan and WinEEG [22] and converted to input data for next phase. WinEEG allows the recording, editing and analysis of continuously recorded EEG and Evented Related Potential data (with the use of PsyTask), and performs database comparisons. Neuroscan provides a set of systems for the acquisition and analysis of EEG and ERP data. The systems are integrated platforms, designed to allow uncompromising solutions for seamless recording and analysis of EEG data across a variety of domains. Neuroscan has developed multiple hardware and software systems that combine to build the ideal platform for a particular area of research.

The second phase is data analysis and decomposition; ERP data contains mixture of brain signals (patterns) and noises. The aim of this phase is to separate signals from noises and disentangle overlapping patterns. [5] There are variety of known tools for data preprocessing and decomposition. For e.g. Net Station [8] is a suite of tools which includes data cleaning, statistical extraction and visualization techniques. EEGLAB [7] is a Matlab toolbox that provides advanced statistical methods for EEG/MEG and ERP processing, with Independent component analysis ICA and Joint time-frequency analysis TFA. Dien PCA toolbox [6] is Principal component analysis PCA tolls that are optimized for ERP data decomposition.

Jia-Cai Zhang and Xiao-Jie Zhao [21] have suggested ICA for decomposition of ERP EEG patterns. "ICA can blindly separate the input ERP data into a sum of temporally independent and spatially fixed components arising from distinct or overlapping brain regions." They used ICA to illustrate "the P300 components in two ERPs recorded under various conditions or tasks. Both are contributed from a few independent sources. ICA decomposition also indicates a new method to compare P300 components between two ERPs induced by two related tasks". After decomposition in this phase there is summary extraction from PCA/ICA components.

The third phase is data mining. Jiawei Rong and Dejing Dou [5] have suggested to split data mining process into two steps; unsupervised and supervised learning. In unsupervised learning; the clustering is applied by using Expectation Maximization EM, it is trying to form the clusters base on summary metrics extracted from e.g PCA, and then the observations that belong to same pattern are expected to map to the same cluster. Later in supervised classification, decision tree is used to generate rules for each cluster and classify them. After classified rules for each cluster is generated, the fourth phase is rule comparison with expert rules. This phase is called evaluation part. Dejing Dou and Gwen Frishkoff [5] have suggested an ERP evaluation phase to refine the rules by ERP expert standards. In this phase the machine generated rules are compared with expert rules and the results would be the ERP domain rules.

ERP sharing is another important issue that concerns experts in this domain. Paea LePendu and Dejing Dou [20] have proposed an automatic method for modeling a relational database that uses SQL triggers and foreign-keys to efficiently answer positive semantic queries about ground instances for Semantic Web ontology. They applied this methodology to the study of brain EEG and ERP data. Dejing Dou [5] has suggested Web ontology Language (WOL) to publish and share ERP domain ontologism on the Web to be accessed by ERP experts.

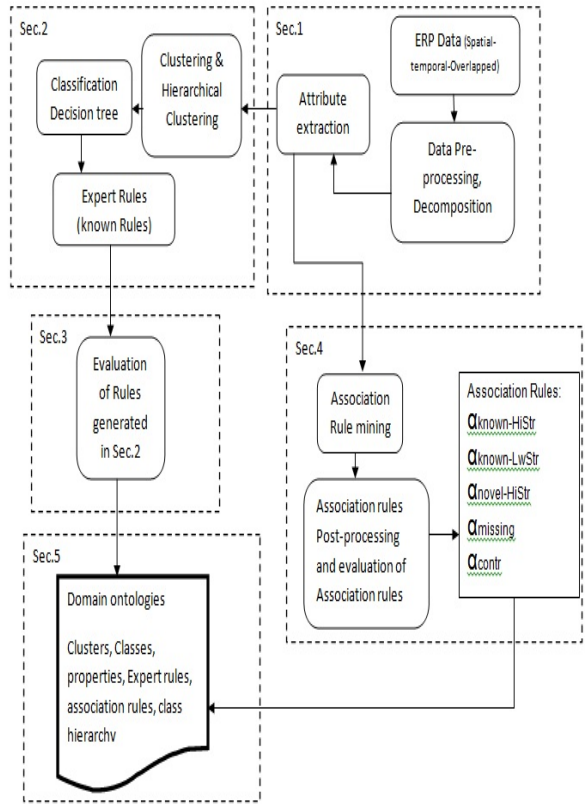

Fig.4. Neuroelectromagnetic Ontology Framework (NOF)

## 3. Neuroelectromagnetic Ontology Framework (NOF)

In Fig.4 there is an illustration of Neuroelectromagnetic Ontology Framework (NOF). first step is data pre-processing following by second step which is data mining. The third step is to evaluate machine generated rule. And finally the last step is association rule mining. The generated rules later will be evaluated with domain ontologies and refined to create the domain hidden rules.

## 4. Data preprocessing

### 4.1 Data Decomposition

ERP data is a mixture of noise which is a combination of artifacts and brain activities that is not related to events, and signals which is the target and the brain pattern. Decomposition methods help detecting noises and separating signals from overlapping patterns. There are many decomposition methods available such as PCA (Principle Component Analysis), Temporal PCA, and ICA (Independent Component Analysis) and Wavelet. In the present paper, ICA is proposed to decompose the data. ICA is a method for separating a multivariate signal into additive subcomponents supposing the mutual statistical independence of the non-Gaussian source signals. The ICA toolbox in Matlab [6] is a useful tool in order to decompose the Matlab-friendly data. The datasets are used as input to the ICA is formed by variables that corresponding to time points as latency in ERP.

### 4.2 ERP Attributes extraction

After Decomposition, a summary of attributes of processed data is extracted. It represents spatial, temporal and functional dimensions of patterns of interest. In this study 13 attributes has been used by adding more spatiotemporal ones. In table 1, there is a list of attributes that are used.

Table 1: shows 13 ERP extracted attributes

| Attribute | Description |
|---|---|
| SP-min(ROI) | channel grouping(ROI) to which the min channel belongs |
| SP-max | channel with max weighting for factor FA |
| SP-max (ROI) | channel grouping(ROI) to which the max channel belongs |
| SP-min | channel with min weighting for factor FA |
| IN-min | min amplitude |
| IN-max | max amplitude |
| IN-mean | mean amplitude for a specified channel set |
| ROI | region of interest |
| SP-cor | cross-correlation between Factor(FA) topography and topography of target pattern |
| TI-max | max latency(time of max amplitude) |
| EVEN | Event type (stimon, respon, EKG-R, etc.) |
| STIM | stimulus |
| MOD | modality of stimulus |

## 5. Data mining processes

### 5.1 Unsupervised learning: ERP Patterns Clustering

Usually, ERP patterns are define through a "manual" process that involves visual assessment of peaked-major-averaged ERP data. Nevertheless, the appropriate definition of a target pattern, its operationalization, and measurement across individual subjects, can vary substantially across research groups. While this method can lead to agreement on the high-level rules and concepts that characterize ERP patterns in a given domain, operationalization of these rules and concepts is highly variable across research labs.

The clustering method that is used in this paper is the PCA Clustering method which is formally explained and used in EEGLAB [7]. Before clustering, we need to prepare the data. This requires, first, identifying the components from each dataset to be entered into the clustering, then computing component activity measures for each study dataset. For this purpose, for each dataset component, the pre-clustering function first computes desired condition-mean measures used to determine the cluster 'distance' of components from each other. The condition means used to construct this overall cluster 'distance' measure may be selected from a palette of standard EEGLAB measures: ERP, power spectrum, ERSP, and/or ITC, as well as the component scalp maps (interpolated to a standard scalp grid) and their equivalent dipole model locations (if any). [7] After clusters are formed up and based on the clustering result, we can generate ontology classes.

### 5.2 Hierarchical Clustering

In this part it is expected to automatically discover the patterns those are found during clustering and arranged hierarchically, then generate taxonomy of the patterns. We may put whole of data in one cluster to sub-divide this cluster into 2 clusters then after that each cluster will be sub-divided repeatedly [9], in each step we take out the majority of data instances of one pattern into one cluster and label it. In another method we may determine the number of clusters based on the previous clustering step and put the data into clusters then try to merge each 2 clusters that are close to each other to form up a higher cluster. This process is done repeatedly until all data instances are put into on cluster. The discovered hierarchy (class taxonomy) can be represented in form of ontology classes and added into the ERP ontology.

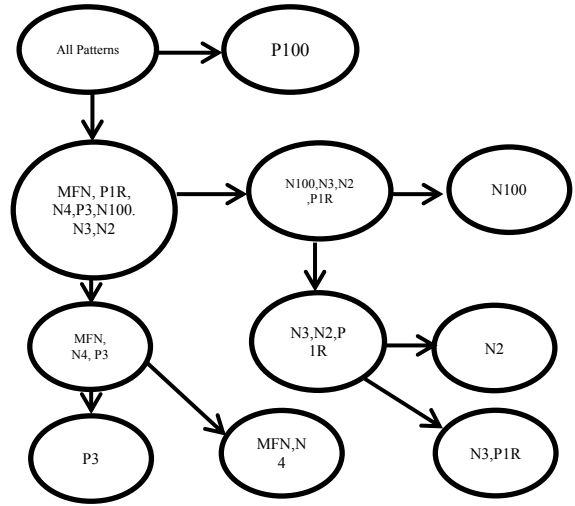

Fig.5 hierarchical clustering [5]

### 5.3 Supervised learning: Cluster-based classification

In clustering, data is automatically partitioned into clusters. The main purpose of classification in this section is to develop rules that accurately assign designated data to clusters. Therefore, we use classification methods such as decision tree to develop mentioned rules. We use C5.0 classification algorithm [16] to classify factors in each cluster. C5.0 is a new decision tree learning algorithm. In other researches C4.5 is mainly used for classification but C5.0 have some significant advantages over C4.5 such as accuracy, speed and memory usage. C5.0 works well for continuous values. Some ERP attributes have continuous values. On the other hand, the classification rules derived from the decision tree are meaningful to human experts. Although data in each cluster can be labeled with cluster names not respective to expert labels. When the clustering process reach to optimized point, then there is no need for the need for expert labeling of patterns, this will be providing a considerable savings in time and a sizable gain in information processing for ERP analysts. The rules will be represented in form of ontology rules to be added into ERP ontology database.

## 6. High-level Knowledge Generation

### 6.1 Rule comparison with domain experts

In decision tree we can generate ERP domain rule that would give a chance to compare them with the domain experts` rules which are manually generated. It is helpful to label the clusters in case their name is arbitrarily assigned. These rules could be a good reference for domain experts.

## 6.2 Association Rule Mining

In clustering and Classification sections the main goal is to distinguish the classes and its properties. Along these classes, the relationship between properties and patterns are important to domain experts and is an input in ERP ontology. Association rule mining finds frequencies among patterns in certain data sets. In this paper, association rule mining is used to seek the properties that frequently co-occur for the specific ERP pattern factors. Values of each attribute are quantified by using their splitting point value in the decision tree. After converting the numeric values of each attribute to categorical values. Then, Apriori algorithm [14] in Weka [13] will be used to find association rules of these attributes.

## 7. Association rules Post-processing

Gunjan Mansingh, Kweku-Muata Osei-Bryson, Han Reichgelt [15] have proposed a generic method for pruning and partitioning the output of association rule induction. This approach can be used to combine the knowledge represented in ERP ontology with an objective measure Reliability (r) to create meaningful partitions in a set of extracted association rules. The partitions that are interesting to the domain experts are; 1. Novel rules with high strengths $\alpha_{novel-HiStr}$ – Extracted association rules those are not part of the ontology and have a Reliability value greater than a user-specified threshold. Rules in this partition may lead to the formation of new beliefs and additions to the ontology. 2. Known rules with high strengths $\alpha_{known-HiStr}$ – Extracted association rules that are already part of the ontology and have a Reliability value greater than a threshold. These rules strengthen previous beliefs of ERP patterns which have been induced by the domain expert(s). 3. Known rules with low strengths $\alpha_{known-LwStr}$ – Extracted association rules that are part of the ontology and have a Reliability value lower than a user-specified threshold. These rules represent aspects of an expert's knowledge that are not supported by data which suggests that the corresponding beliefs in the ERP domain that may no longer be true. When such rules are extracted the domain experts may investigate the reasons behind such an occurrence. 4. Missing rules $\alpha_{missing}$ – Expert Rules in ERP domain were not extracted by association rule induction. Presence of these rules implies that either the threshold values are incorrect or that the domain knowledge is not supported by data. 5. Contradictory rules $\alpha_{contr}$ – Rules extracted by association rule induction with high strengths that contradict Expert rules. In order to detect the presence of such rules, the representation language in which the rules and the original ontologies were expressed must allow for the expression of negations.

In order to prune and create partitions in the output of the association rule induction process we use the domain ontology and reliability measures. the thresholds should be determined by experts; $\beta_{sup}$ (minimum support), $\beta_{conf}$ (minimum confidence) and $\pi_{min}$ (minimum reliability). We extract the previously known rules $\alpha_{pre-known}$ (or expert rules) from domain ontology. Then we constrain the association rules by Support and Confidence thresholds $\beta_{sup}$ $\beta_{conf}$ therefore a new set of association rules based on expert constrains $\alpha AREC$ are represented then for each $\alpha_{AREC}$, the $\pi_{rel}$ (reliability) need to be calculated.

Fourth the definitions of rules that are found in the process are below.

| |
|---|
| $\alpha_{known-HiStr} = \alpha_{AREC} \cap \alpha_{pre-known} \ \{\pi_{rel}, \alpha_{AREC} >= \pi_{min}\}$ |
| $\alpha_{known-LwStr} = \alpha_{AREC} \cap \alpha_{pre-known} \ \{\pi_{rel}, \alpha_{AREC} < \pi_{min}\}$ |
| $\alpha_{novel-HiStr} = \alpha_{AREC} - \alpha_{pre-known} \ \{\pi_{rel}, \alpha_{AREC} >= \pi_{min}\}$ |
| $\alpha_{missing} = \alpha_{pre-known} - \alpha_{AREC}$ |
| $\alpha_{contr} = \{ A \in \alpha_{AREC} \mid B \in \alpha_{pre-known} : A = -B \}$ |

## 8. Conclusion

We gave a brief overview of ERP domain knowledge in MEG EEG brain signals, and introduce Neuroelectromagnetic Ontology Framework (NOF) for mining event-related potentials as well as, data-mining process and results. The aim for this paper is to develop an infrastructure for mining, analysis and sharing the ERP domain ontologies in EEG, MEG data. The proposed Neuroelectromagnetic Ontology Framework (NOF) for mining ERP patterns contains 5 stages: 1) Data pre-processing and preparation: in this phase we collect brain signals by Netstation software and filter the noises along with decomposition of overlapping signals ; 2) Data mining application: in this phase we apply data mining which include unsupervised and supervised methods; 3) Rule Comparison and Evaluation: in this phase we refine the expert rules and evaluate the machine generated rules; 4) Association rules Post-processing: in this phase we do association rule mining and refine the association rules by domain ontologies to produce new set of hidden rules. 5) Domain Ontologies: in this phase we develop the ERP ontologies from the results of previous phases. The domain ontologies are ERP inside knowledge and will be stored in a knowledge base. The results shows machine generated rules are good reference for expert rules and these rules can be evaluated by rules from Decision tree and Association rule mining. On the other hand, in Association rules Post-processing a new set of hidden rules can be discovered based on association rules and domain ontologies to improve and refine expert rules. The outcome of this research is a Neuroelectromagnetic knowledge-based system which contains the ERP domain ontologies. This System can refine the existing expert knowledge and produce ERP inside knowledge which will be reference for experts in the field of neuroscience. This knowledge extends the expert understanding of brain functions and will be used in finding the causes of brain disorders Cortex in-functionality, Neuron illness, brain capability analysis, and responses to outside stimulus in different situations. Although it can be used in commercial marketing known as Neural-Economy to increase sale and find the profitable customers.

**Acknowledgments**

This work was supported by a grant from the University Sains Malaysia. We thank Professor Jafri Malin Abdullah Department of Neuroscience, School of Medical Sciences and Prof. Rahmat Budiarto for helping us in this paper.

**References**

[1] Niedermeyer E. and da Silva F.L. (2004). Electroencephalography: Basic Principles, Clinical Applications, and Related Fields. Lippincot Williams & Wilkins.

[2] Cohen D. "Magnetoencephalography: evidence of magnetic fields produced by alpha rhythm currents." Science 1968; Vol.5, No.6, pp.161-784

[3] Coles, Michael G.H.; Michael D. Rugg (1996). "Event-related brain potentials: an introduction". Electrophysiology of Mind. Oxford Scholarship Online Monographs. pp. 1–27.

[4] Kam Swee Nga, , Hyung-Jeong YangCorresponding , Sun-Hee Kima, "Hidden pattern discovery on event related potential EEG signals". Department of Computer Science, Chonnam National University, South Korea, Vol. 2, No.1, PP. 123-129.

[5] Rong, J., Dou, D., Frishkoff, G., Frank, R., Malony, A., & Tucker, D. (2007). A semi-automatic framework for mining ERP patterns. Proceedings of the IEEE International Symposium on Data Mining and Information Retrieval (DMIR-07), Ontario, Canada. pp. 329-334.

[6] Dien, J. PCA Toolbox (version 1.7).

[7] EEG lab. http://www.sccn.ucsd.edu/eeglab/.

[8] NetStation Technical Manual. http://www.egi.com.

[9] Frishkoff, G., Frank, R., Rong, J., Dou, D., Dien, J., & Halderman, L. (2007). A framework to support automated ERP pattern classification and labeling. Computational Intelligence and Neuroscience, Vol.1, No.3, 2007, Article ID 14567.

[10] LePendu, P., Dou, D., Frishkoff, G. , & Rong, J. (2008). Semantic data modeling: Methods for ontology-based queries and an application to brainwave data., Proceedings of the 20th International Conference on Scientific and Statistical Database Management (SSDBM-08), Hong Kong, China.Vol.20, No.2, pp.220-228.

[11] Dou, D., Frishkoff, G., Rong, J., Frank, R., Malony, A., & Tucker, D. (2007). Development of NeuroElectroMagnetic Ontologies (NEMO): A framework for mining brain wave ontologies, Proceedings of the Thirteenth International Conference on Knowledge Discovery and Data Mining (KDD2007), pp. 270-279, San Jose, CA.

[12] Christos N. Zigkolis, Nikolaos A. Laskaris, "Using conditional FCMtomine event-related brain dynamics" Journal of Computers in Biology and Medicine 39 (2009) 346 – 354, Artificial Intelligence & Information Analysis Laboratory, Department of Informatics, Aristotle University, Thessaloniki, Greec

[13] Weka 3: Data Mining Software in Java. http://www.cs.waikato.ac.nz/ml/weka/.

[14] R. Agrawal and R. Srikant. Fast Algorithms for Mining Association Rules. In Very Large Data Bases (VLDB) Conference, pages 487–499. Morgan Kaufmann, 1994.

[15] Gunjan Mansingh, Kweku-Muata Osei-Bryson, Han Reichgelt. "Using ontologies to facilitate post-processing of association rules by domain experts" International Journal of Information Sciences 181 (2011) 419–434.

[16] Ron Kohavi, Ross Quinlan. "Decision Tree Discovery" robotics.stanford.edu

[17] Johnjoe McFadden (2002). "The Conscious Electromagnetic Information (Cemi) Field Theory: The Hard Problem Made Easy?". Journal of Consciousness Studies. Vol. 9, No.8, pp 45–60.

[18] http://sccn.ucsd.edu/~arno/fam2data/publicly_available_EEG_data.html.

[18] http://sccn.ucsd.edu/~arno/fam2data/publicly_available_EEG_data.html.

[19] M. Sabeti, S.D. Katebi, R. Boostani, G.W. Price, "A New Approach for EEG Signal Classification of Schizophrenic and Control Participants" Elsevier Journal of Expert Systems with Applications, Volume 38, Issue 3, March 2011, Pages 2063-2071 (ISI)

[20] Paea LePendu, Dejing Dou, Gwen A. Frishkoff, Jiawei Rong: Ontology Database: A New Method for Semantic Modeling and an Application to Brainwave Data. SSDBM 2008: 313-330 [22] C. Davatzikos, M. Acharyya, K. Ruparel, D.G. Shen, J. Loughhead, R.C. Gur, and D. Langleben , "Classifying spatial patterns of brain activity for lie-detection", NeuroImage, 28 (3), 663-668, 2005.

[21] Jia-Cai Zhang, Xiao-jie Zhao, Yi-Jun Liu, Li Yao: Mining the Independent Source of ERP Components with ICA Decomposition. ISNN (2) 2006: 592-599

[22] http://www.novatecheeg.com/wineeg.html

**First Author** Seyed Aliakbar Mousavi Aslarzanagh is a researcher in University Science Malaysia, School of Computer Science. He obtained E-commerce Bachelor degree from Multimedia University in 2010 and continued his study to a research degree in Computer Science. He is currently cooperating with Department of Neuroscience in USM Kubang Kerian and multimedia research group in an interdisciplinary fields of Neuroinformatic and Data Mining Knowledge discovery.

**Second Author** Assoc Prof Muhammad Rafie Hj. Mohd. Arshad achieved his Bachelor degree from Macalester College, Minnesota, U.S.A. and his MBA-MIS from Dallas, Texas, U.S.A.

**Third Author** Hasimah Hj Mohamed has obtained her bachelor degree from Universiti Teknology Malaysia and her MSc from Universiti Sains Malaysia. Currently she is a senior lecturer at Universiti Sains Malaysia, and also a PhD candidate at National University of Malaysia.

**Fourth Author** Saleh Ali K. Al-Omari has obtained his Bachelor degree in Computer Science from Jerash University, Jordan in 2004-2005 and Master degree in Computer Science from Universiti Sains Malaysia, Penang, Malaysia in 2007. Currently, He is a PhD candidate at the School of Computer Science, Universiti Sains Malaysia. He is the candidate of the Multimedia Computing Research Group, School of Computer Science, USM. He is a member and reviewer of several international journals and conferences (IEICE, JDCTA, IEEE, IACSIT, AIRCC, etc). His main research area interest are in includes Peer to Peer Media Streaming, Video on Demand over Ad Hoc Networks, MANETs, and Multimedia Networking, Mobility.